\documentclass[amsthm]{elsart}
 \pdfoutput=1
\usepackage{yjsco}
\usepackage{natbib}
\usepackage{algorithmicx}
\usepackage{algcompatible}
\usepackage{amsmath}
\usepackage{amscd}
\usepackage{amssymb}
\usepackage{amsfonts}
\usepackage{indentfirst}
\usepackage{url}

\def \R {{\mathbb R}}

\def  \RealTriangularize {{\tt RealTriangularize}}

\def \Discover {{\tt Discover}}

\def \sign {{\rm sign}}

\begin{document}

\begin{frontmatter}
\title{\textbf{A Simple Quantifier-free Formula of Positive Semidefinite Cyclic Ternary Quartic Forms}}
\author{Jingjun Han\corauthref{cor}}
\corauth[cor]{Corresponding author.}
\ead{hanjingjunfdfz@gmail.com}
\address{School of Mathematical Sciences, Peking University, Beijing 100871, China}
\thanks{This version will appear in Proc. the 10th Asian Symposium on Computer Mathematics (ASCM 2012).}
\begin{abstract}
 The problem of quantifier elimination of positive semidefinite cyclic ternary quartic forms is studied in this paper. We solve the problem by function \RealTriangularize\ in Maple15, the theory of {\em complete discrimination systems} and the so-called {\em Criterions on Equality of Symmetric Inequalities method}. The equivalent simple quantifier-free formula is proposed, which is difficult to obtain automatically by previous methods or quantifier elimination tools.
\end{abstract}
\end{frontmatter}
\section{Introduction}
The elementary theory of real closed fields (RCF) is expressed in a formal language with atomic
formulas of the forms $A = B$ and $A > B$, where $A$ and $B$ are multivariate
polynomials with integer coefficients. The problem of quantifier elimination (QE) for RCF can be expressed as: for a given formula of RCF, find an equivalent formula containing the same
free (unquantified) variables and no quantifiers.  \par
QE problem is what many researchers have contributed to, including A. Tarski, who gave a first quantifier
elimination method for real closed fields in the 1930s, although its publishing delayed for nearly 20 years \citep{Ta48}, and G. E. Collins, who introduced a so-called cylindrical algebraic decomposition (CAD) algorithm for QE problem in the 1970s \citep{Co75}, which has turned into one of the main tools for QE problems, along with its improved variations. Over the years, new algorithm and important improvements on CAD have appeared, including, for instance, \citep{ACM84b,ACM88,Mc88,Hong90,CH91,Hong92}
and \citep{Co98,Mc98,Wei98,Br01a,Br01b,BM05,MB09,Br12}. Most of the works, including Tarski¡¯s algorithm, were collected in a book \citep{CJ98}.\par
Many researchers have studied a special quantifier elimination problem (see,
for example, \citep{AM88,La88,CH91,Wei94}),
$$(\forall x\in \R)(x^4 + px^2 + qx + r \ge 0),$$
which is called {\em quartic problem} in the book just mentioned.
There are also many researchers that have studied some special QE problems in other ways. In 1987, Choi etc. obtained the necessary and sufficient condition for the positive semidefiniteness of a symmetric form of degree 3 with $n$ variables \citep{CLR87}. Gonz\'{a}lez-Vega etc. proposed a theory on root classification of polynomials in \citep{GLRR89} which is based on the Sturm-Habicht sequence and the theory of subresultants. For QE problems in the form $(\forall x)(f(x) \ge 0)$ where the degree of $f(x)$ is a positive even integer, Gonz\'{a}lez-Vega proposed a combinatorial algorithm \citep{Gon98} based on the work in \citep{GLRR89}. In 1996, Yang etc. proposed the theory of {\em complete discrimination systems} for polynomials to discuss the root classification problem of one variable polynomial with real parameters \citep{YHZ96,Yang99a}. Yang's theory is equivalent to Gonz\'{a}lez-Vega's. In 1999, Harris gave a necessary and sufficient condition for the positive semidefiniteness of a symmetric form of degree 4 and 5 with 3 variables \citep{Ha99}. In 2003, Timofte considered the necessary and sufficient condition for the positive semidefiniteness for symmetric forms of degree $d$ with $n$ variables in
$\R^n$$(d\le5)$ \citep{Ti03,Ti05}. By applying Timofte's result and the theory of {\em complete discrimination systems}, Yao etc. obtained a quantifier elimination of the positive semidefiniteness for symmetric forms of
degree $d$ with $n$ variables in $\R^n$$(d\le5)$ \citep{YF08}. However, the above results are for symmetric forms. Therefore, the author discussed the positive semidefiniteness for more general forms with $n$ variables, including symmetric forms and cyclic forms \citep{Han11}. \par
In this paper, we consider a quantifier-free formula of positive semidefinite cyclic ternary quartic forms, namely the quantifier-free formula of
$$(\forall x,y,z\in \R)[F(x,y,z)=\sum_{cyc}x^4+k\sum_{cyc}x^2y^2+l\sum_{cyc}x^2yz+m\sum_{cyc}x^3y+n\sum_{cyc}xy^3\ge0],$$
which is similar to yet also more complex than {\em quartic problem}. It is difficult to get an answer directly by previous methods or QE tools. Recall Hilbert's 1888 theorem that says, every positive semidefinite ternary quartic (homogeneous
polynomial of degree 4 in 3 variables) is a sum of three squares
of quadratic forms \citep{Hilbert}. Hilbert's proof is non-constructive in the sense that it
gives no information about the production of an equivalent quantifier-free formula.
 Notice that $F(x,y,z)\ge0$ for $x,y,z\in \R$ is equivalent to the following inequality $$(\forall x,y,z\in \R)[ f(x,y,z)=\sigma_1^4+B\sigma_1^2\sigma_2+C\sigma_2^2+D\sigma_1\sigma_3+E\sigma_1\sum_{cyc}{x^2y}\ge0],$$
where $\sigma_1=x+y+z$, $\sigma_2=xy+yz+zx$, $\sigma_3=xyz$ and $B,C,D,E$ satisfying
$$k=2B + C + E + 6,l=2C+D+E+12+5B,m=B+4,n=B+E+4.$$
The author \citep{Han11} obtained the following necessary and sufficient condition of $f(x,y,z)\ge0$,
$$(\forall m\in \R)[f(m,1,km+1-k)\ge0],$$
where $k$ is a real root of the equation
$$Ek^3-Dk^2-3Ek+Dk+E=0.$$
 However, it is still difficult to get a quantifier-free formula by previous methods or QE tools.\par
The author developed several other methods to solve cyclic and symmetric inequalities including the so-called {\em Criterions on Equality of Symmetric Inequalities method} \citep{Han11}. These methods can solve a class of QE problems. This paper is firmly rooted in the author's book \citep{Han11}, especially the technique dealing with the cyclic and symmetric inequalities. In order to be self contained, we will prove some results later in this paper. In order to obtain a simple quantifier-free formula, function \RealTriangularize \citep{CDMMXX10} of RegularChains package in Maple15 is used to prove inequalities and illustrate semi-algebraic systems
without real solution. We also need the theory of {\em complete discrimination systems} for root
classification.\par
The rest of the paper is organized as follows. Section 2 introduces some basic
concepts and results about {\em complete discrimination systems} for polynomials. Section 3 presents our solution to the positive semidefinite cyclic ternary quartic form.
\section{Preliminaries}
Given a polynomial
$$f(x)=a_0x^n+a_1x^{n-1}+\cdots+a_n,$$
we write the derivative of $f(x)$ as
$$f'(x)=0\cdot x^n+na_0x^{n-1}+(n-1)a_1x^{n-2}+\cdots+a_{n-1}.$$
\begin{defn}\citep{YHZ96,Yang99a} ({\em discriminant matrix})
The Sylvester matrix of $f(x)$ and $f'(x)$
\[\left|
\begin{matrix}
   &a_0&a_1&a_2&\ldots  &a_n&&&  \\
   &0&na_0&(n-1)a_1&\ldots &a_{n-1} &&&\\
   &&a_0&a_1&\ldots  &a_{n-1}&a_n&&\\
   &&0&na_0&\ldots &2a_{n-1} &a_n&&\\
   &                &\vdots                 &\vdots  & \ddots & \vdots &\vdots  \\
   &&&&&a_0&a_1&\ldots&a_{n}\\
   &&&&&0&na_0&\ldots&a_{n-1}
\end{matrix}\right|\]
\end{defn}
is called the {\em discrimination matrix} of $f(x)$, and denoted by $Discr(f)$.
\begin{defn}\citep{YHZ96,Yang99a} ({\em discriminant sequence})
Denoted by $D_k$ the determinant of the submatrix of $Discr(f)$ formed by the first $2k$ rows and the first $2k$
columns. For $k=1,\ldots,n$, we call the $n$-tuple
$$\{D_1(f),D_2(f),\ldots,D_n(f)\}$$
the {\em discriminant sequence} of polynomial $f(x)$.
\end{defn}
\begin{defn}\citep{YHZ96,Yang99a} ({\em sign list}). We call list
$$[\sign(D_1(f)), \sign(D_2(f)), \ldots, \sign(D_n(f))]$$
the {\em sign list} of the {\em discriminant sequence}$\{D_1(f),D_2(f),\ldots,D_n(f)\}$
\end{defn}
\begin{defn}\citep{YHZ96,Yang99a} ({\em revised sign list}).
Given a {\em sign list} $$[s_1,s_2,\ldots,s_n],$$
we construct a new list
$$[\epsilon_{1},\epsilon_{2},\ldots,\epsilon_{n}]$$
as follows (which is called the {\em revised sign list}):
if $[s_1,s_2,\ldots,s_n]$ is a section of the give list, where
$s_i\neq0,$ $s_{i+1}=s_{i+2}=\ldots=s_{i+j-1}=0,$ $s_{i+j}\neq0,$ then we replace the subsection
$$[s_{i+1},s_{i+2},\ldots,s_{i+j-1}]$$
by
$$[-s_i,-s_i,s_i,s_i,-s_i,-s_i,s_i,s_i,\ldots],$$
i.e., let
$$\epsilon_{i+r}=(-1)^{[\frac{r+1}{2}]}\cdot s_i$$
for $r=1,2,\ldots,j-1.$ Otherwise, let $\epsilon_{k}=s_k$ i.e., no change for other terms.
\end{defn}
\begin{lem}\citep{YHZ96,Yang99a}\label{thm1}
Given a polynomial with real coefficients,
$f(x)=a_0x^n+a_1x^{n-1}+\cdots+a_n.$
If the number of the sign changes of the {\em revised sign list} of
$$\{D_1(f),D_2(f),\ldots,D_n(f)\}$$
is $v$, then the number of the pairs of distinct conjugate imaginary root of $f(x)$ equals $v$.
Furthermore, if the number of non-vanishing members of the {\em revised sign list} is $l$, then the
number of the distinct real roots of $f(x)$ equals $l-2v$.
\end{lem}
Theoretically, we can get a quantifier-free formula of the positive semidefinite cyclic ternary quartic form by {\em complete discrimination
systems for polynomials}. But it is impossible because of the complexity.
\section{Main result}
\begin{lem}\label{lem:range}\citep{Han11}
Let $x,y,z\in \mathbb{C}$, $x+y+z=1$ and $xy+yz+zx,xyz\in \R$. The necessary and sufficient condition of $x,y,z\in \R$ is $xyz\in[r_1,r_2]$, where $$r_1=\frac{1}{27}(1-3t^2-2t^3),r_2=\frac{1}{27}(1-3t^2+2t^3)$$ and $t=\sqrt{1-3(xy+yz+zx)}\ge0$.
\begin{pf}
  We consider the polynomial
  $$f(X)=X^3-(x+y+z)X^2+(xy+yz+zx)X-xyz,$$
  it is obvious that $x,y,z$ are three roots of $f(X)=0$.
  By Lemma \ref{thm1}, the equation $f(X)=0$ has three real roots if and only if
  $$D_3(f)\ge0 \wedge D_2(f)\ge0,$$
  where
 $$D_2(f)=(x+y+z)^2-3(xy+yz+zx)=1-3(xy+yz+zx),$$ $$D_3(f)=(x-y)^2(y-z)^2(z-x)^2=\frac{1}{27}(4D_2(f)^3-(3D_2(f)-1+27xyz)^2).$$
 Therefore, using the substitution $t=\sqrt{D_2(f)}$ and $xyz=r$, we have
    $$x,y,z\in \R \Longleftrightarrow (x-y)^2(y-z)^2(z-x)^2\ge0 \wedge (x+y+z)^2\ge 3xy+3yz+3zx,$$
     $$\Longleftrightarrow 4t^6-(3t^2-1+27r)^2\ge0 \wedge t\ge0 $$
     $$\Longleftrightarrow \frac{1}{27}(1-3t^2-2t^3)\le r\le \frac{1}{27}(1-3t^2+2t^3) \wedge  t\ge0.$$
        That completes the proof.
       \end{pf}
\end{lem}
\begin{rem}
 The author also get this result by the {\em Criterions on Equality of Symmetric Inequalities method} \citep{Han11}. This Lemma implies that if $x,y,z\in\R$ and $x+y+z=1$, then $\sqrt{1-3(xy+yz+zx)}=t\ge0$, and the range of $xyz$ is $[r_1,r_2]$.
\end{rem}
We now try to reduce the number of quantifiers of the positive semidefinite cyclic ternary quartic form which is mentioned in the Introduction,
$$(\forall x,y,z\in \R)[F(x,y,z)=\sum_{cyc}x^4+k\sum_{cyc}x^2y^2+l\sum_{cyc}x^2yz+m\sum_{cyc}x^3y+n\sum_{cyc}xy^3\ge0].$$
\begin{lem}\citep{Han11}\label{lem:eqv}
 The inequality $F(x,y,z)\ge0$ holds for any $x,y,z\in \R$ if and only if
 \begin{align*}
 &2\sum_{cyc}x^4+2k\sum_{cyc}x^2y^2+2l\sum_{cyc}x^2yz+(n+m)\sum_{cyc}x^3y+(m+n)\sum_{cyc}xy^3\\
 \ge&|(m-n)(x+y+z)(x-y)(y-z)(z-x)|
 \end{align*}
 holds for all $x,y,z\in \R$.
 \begin{pf}
   It is easy to show that for all $x,y,z\in \R$, $F(x,y,z)\ge0$ is equivalent to$:$ for all $x,y,z\in\R$,
   \begin{align*}
   &2\sum_{cyc}x^4+2k\sum_{cyc}x^2y^2+2l\sum_{cyc}x^2yz+(n+m)\sum_{cyc}x^3y+(m+n)\sum_{cyc}xy^3\\
 \ge&(m-n)(x+y+z)(x-y)(y-z)(z-x).
 \end{align*}
 On the other hand, if $F(x,y,z)\ge0$ holds for any $x,y,z\in \R$, then $F(x,z,y)\ge0$ also holds for any
 $x,y,z\in\R$. This inequality is equivalent to
   \begin{align*} &2\sum_{cyc}x^4+2k\sum_{cyc}x^2y^2+2l\sum_{cyc}x^2yz+(n+m)\sum_{cyc}x^3y+(m+n)\sum_{cyc}xy^3\\
 \ge&(n-m)(x+y+z)(x-y)(y-z)(z-x)
 \end{align*}
 for all $x,y,z\in\R$.\par
 Thus, $F(x,y,z)\ge0$ for any $x,y,z\in \R$ is equivalent to
  \begin{align*}
  &2\sum_{cyc}x^4+2k\sum_{cyc}x^2y^2+2l\sum_{cyc}x^2yz+(n+m)\sum_{cyc}x^3y+(m+n)\sum_{cyc}xy^3\\
 \ge&|(m-n)(x+y+z)(x-y)(y-z)(z-x)|
 \end{align*}
 for all $x,y,z\in\R$.
 \end{pf}
\end{lem}
\begin{thm}\label{thm:eqv}
The positive semidefinite cyclic ternary quartic form
$$\forall x,y,z\in \R \qquad F(x,y,z)\ge0$$
holds if and only if the following inequality holds.
\begin{align*}(\forall t\in \R)[g(t):=&3(2+k-m-n)t^4+3(4+m+n-l)t^2+k+1+m+n+l-\\
  &\sqrt{27(m-n)^2+{(4k+m+n-8-2l)}^2}t^3\ge0].
  \end{align*}
  \begin{pf}
  Since $\sqrt{27(m-n)^2+{(4k+m+n-8-2l)}^2}t^3\le0$ when $t\le0$, $(\forall t\ge0)[g(t)\ge0]$ implies $(\forall t\in \R)[g(t)\ge0]$, thus they are equivalent. We only need to prove $(\forall x,y,z\in \R)[ F(x,y,z)\ge0]$ is equivalent to $(\forall t\ge0)[g(t)\ge0]$.\par
  According to Lemma \ref{lem:eqv}, the positive semidefinite cyclic ternary quartic form is equivalent to
    \begin{align*}
   &2\sum_{cyc}x^4+2k\sum_{cyc}x^2y^2+2l\sum_{cyc}x^2yz+(n+m)\sum_{cyc}x^3y+(m+n)\sum_{cyc}xy^3\\
 &\ge|(m-n)(x+y+z)(x-y)(y-z)(z-x)|
 \end{align*}
 for all $x,y,z\in\R$.\par
   Substituting $x+y+z,xy+yz+zx,xyz$ with $p,q,r$, we have
   \begin{align*}
   &\sum_{cyc}x^4=p^4-4p^2q+2q^2+4pr \qquad  \sum_{cyc}x^2y^2=q^2-2pr\\
   &\sum_{cyc}x^2yz=pr   \qquad \sum_{cyc}{x^3y+xy^3}=q(p^2-2q)-pr\\
   &|(x-y)(y-z)(z-x)|=\sqrt{(x-y)^2(y-z)^2(z-x)^2}=\sqrt{\frac{4(p^2-3q)^3-(2p^3-9pq+27r)^2}{27}}.
   \end{align*}
      The last inequality above becomes
      \begin{align*}
      G(x,y,z)=&2p^4+np^2q-8p^2q+mp^2q+2kq^2-2nq^2-2mq^2+4q^2+2lpr+8pr-npr-mpr\\
      &-4kpr-|m-n|p\sqrt{\frac{4(p^2-3q)^3-(2p^3-9pq+27r)^2}{27}}\ge0
      \end{align*}
   We first prove the sufficiency.\\
   If $p=0$, then the inequality $G(x,y,z)\ge0$ becomes
   $$2(2+k-m-n)q^2\ge0.$$
   We can deduce $(2+k-m-n)\ge0$ from $g(t)\ge0$ for all $t\ge0$. $($Since $(2+k-m-n)$ is the {\em leading coefficient} of $g(t)$.$)$\\
   If $p\neq0$, since the inequality is homogenous, we can assume that $p=1$. Notice that
   $$(x+y+z)^2\ge 3(xy+yz+zx),$$
   thus we have $q\le \frac{1}{3}$.
   Using the substitution $t=\sqrt{1-3q}\ge0$, the inequality $G(x,y,z)\ge0$ is equivalent to
   \begin{equation}\label{eqn:eqv2}
   \begin{split}
   &2(2+k-m-n)t^4+(16-4k+m+n)t^2-2+2k+m+n+9(8-4k+2l-m-n)r\\
   \ge &\sqrt{3}|m-n|\sqrt{4t^6-(3t^2-1+27r)^2},
   \end{split}
   \end{equation}
   where $t\ge0$, $r\in[r_1,r_2]$ ($r_1$ and $r_2$ are the same as those in the Lemma \ref{lem:eqv}).
   Since $\frac{2}{3}g(t)\ge0$ is equivalent to
   \begin{align*}
     &2(2+k-m-n)t^4+(16-4k+m+n)t^2-2+2k+m+n+9(8-4k+2l-m-n)r\\
     \ge&\frac{2\sqrt{27(m-n)^2+(8-4k+2l-m-n)^2}t^3}{3}+\frac{(8-4k+2l-m-n)(3t^2-1+27r)}{3},
   \end{align*}
   thus, in order to prove $G(x,y,z)\ge0$, it is sufficient to prove that
   \begin{equation}\label{eqn:eqv1}
   \begin{split}
   &\sqrt{3}|m-n|\sqrt{4t^6-(3t^2-1+27r)^2}\\
   \le &\frac{2\sqrt{27(m-n)^2+(8-4k+2l-m-n)^2}t^3}{3}+\frac{(8-4k+2l-m-n)(3t^2-1+27r)}{3}
   \end{split}
   \end{equation}
      After we square both sides and collect terms, the above inequality is equivalent to
   $$H^2(r)\ge0,$$
   where $$H(r)=\frac{2(8-4k+2l-m-n)}{3}t^3+(3t^2-1+27r)\sqrt{3(m-n)^2+\frac{(8-4k+2l-m-n)^2}{9}}.$$
   It is obviously true.\par
   So the sufficiency is proved. Now we prove the necessity, which is equivalent of proving that when the inequality (\ref{eqn:eqv2}) holds for all $x,y,z\in \R$, then $(\forall t\ge0)[g(t)\ge0]$. For any $t\ge0$, if there exist $x,y,z\in \R$ such that $H(r)=0$, $x+y+z=1$ and $1-3(xy+yz+zx)=t^2$, then the equation of inequality (\ref{eqn:eqv1}) could be attained. Choosing such $x,y,z\in \R$, inequality (\ref{eqn:eqv2}) becomes
     \begin{align*}
     &2(2+k-m-n)t^4+(16-4k+m+n)t^2-2+2k+m+n+9(8-4k+2l-m-n)r\ge \\ &\frac{2\sqrt{27(m-n)^2+(8-4k+2l-m-n)^2}t^3}{3}+\frac{(8-4k+2l-m-n)(3t^2-1+27r)}{3},
     \end{align*}
     which is equivalent to $(\forall t\ge0)[g(t)\ge0]$. Thus, it suffices to show that there exist such $x,y,z\in \R$. 
     Notice that
     \begin{align*}
       H(r_1)H(r_2)=&(\frac{2(8-4k+2l-m-n)}{3}t^3-2t^3\sqrt{3(m-n)^2+\frac{(8-4k+2l-m-n)^2}{9}})\\
       &(\frac{2(8-4k+2l-m-n)}{3}t^3+2t^3\sqrt{3(m-n)^2+\frac{(8-4k+2l-m-n)^2}{9}})\\
       =&-12t^6(m-n)^2\le0,
     \end{align*}
     where
     $$r_1=\frac{1}{27}(1-3t^2-2t^3),r_2=\frac{1}{27}(1-3t^2+2t^3).$$
     Therefore, for any given $t=\sqrt{1-3(xy+yz+zx)}\ge0$, there exists $r_0\in[r_1,r_2],$ such that $H(r_0)=0$. By Lemma \ref{lem:range}, such $x,y,z\in \R$ exist and we prove the necessity.\par
     From the above discussion, the theorem is proved.
     \end{pf}
\end{thm}
We will apply function \RealTriangularize\ of RegularChains package in Maple15 to prove the following lemma.
\begin{lem}\label{lem:3}
Let $a_0>0$, $a_4>0$, $a_1\neq0$, $a_1, a_2\in \R$, we consider the following polynomial
$$f(x)=a_0x^4+a_1x^3+a_2x^2+a_4.$$
The {\em discriminant sequence} of $f(x)$ is
$$D_f=[D_1(f), D_2(f), D_3(f), D_4(f)],$$
where
\begin{align*}
D_1(f)=&{a_0}^2,\\
D_2(f)=&-8a_0^3a_2+3a_1^2a_0^2,\\
D_3(f)=&-4a_0^3a_2^3+16a_0^4a_2a_4+a_0^2a_1^2a_2^2-6a_0^3a_1^2a_4,\\
D_4(f)=&-27a_0^2a_1^4a_4^2+16a_0^3a_2^4a_4-128a_0^4a_2^2a_4^2-\\
&4a_0^2a_1^2a_2^3a_4+144a_0^3a_2a_1^2a_4^2+256a_0^5a_4^3.
\end{align*}
For all $x\in \R$, $f(x)\ge0$ holds if and only if one of the following cases holds,
\begin{align*}
&(1)D_4(f)>0\wedge(D_2(f)<0\vee D_3(f)<0),\\
&(2)D_4(f)=0,D_3(f)<0.
\end{align*}
\begin{pf}
$\Longrightarrow:$ If $f(x)\ge0$ holds for all $x\in \R$, then the number of distinct real roots of $f(x)$ is less than $2$. If it equals $2$, then the roots of $f(x)$ are all real. If it equals $0$, then $f(x)$ has no real root.\par
If $D_4(f)<0$ and $D_2(f)>0$, then the number of non-vanishing members of {\em revised sign list}, $l$, equals $4$. Since $D_4(f)D_2(f)<0$, then the number of the sign changes of {\em revised sign list}, $v$, equals $1$, thus $l-2v=2$. By Lemma \ref{thm1}, the number of distinct real roots of $f(x)$ equals two and the number of the pairs of distinct conjugate imaginary root of $f(x)$, $v=1$, which is impossible. Using function \RealTriangularize, we can prove that the semi-algebraic system $a_4>0, D_4(f)<0,D_2(f)\le0$ has no real solution. Therefore, $D_4(f)\ge0$. Since $D_1(f)\ge0$, the number of the sign changes of {\em revised sign list} $v\le2$.\par
If $D_4(f)>0$, thus $l=4$. Notice that the number of real roots of $f(x)$, namely $l-2v\le2$, so $v\ge1$, from which, we get
$$D_2(f)\le0\vee D_3(f)\le0.$$
Using function \RealTriangularize, we can prove that both the semi-algebraic system $a_4>0,a_0>0,D_4(f)>0,D_2(f)\ge0,D_3(f)=0,a_1\neq0$ and the semi-algebraic system $a_4>0,a_0>0,D_4(f)>0,D_3(f)\ge0,D_2(f)=0,a_1\neq0$ have no real solution.
Hence, if $D_4(f)>0$ and $D_2(f)=0$, then $D_3(f)<0$; if $D_4(f)>0$ and $D_3(f)=0$, then $D_2(f)<0$. Thus, when $D_4(f)>0$, either $D_2(f)<0$ or $D_3(f)<0$ holds. \par
If $D_4(f)=0$ and $D_3(f)>0$, then $l=3$. The number of sign changes of {\em revised sign list} $v$ equals either
$2$ or $0$. From $0\le l-2v\le2$, we have $v=1$, which leads to contradiction. That implies if $D_4(f)=0$, then $D_3(f)\le0$.
Using function \RealTriangularize, we can prove that the semi-algebraic system $a_4>0,a_0>0,D_4(f)=0,D_3(f)=0,a_1\neq0$ has no real solution. Hence, when $D_4(f)=0$, we have $D_3(f)<0.$ \par
$\Longleftarrow:$ If $D_4(f)>0\wedge(D_2(f)<0\vee D_3(f)<0)$, then the number of sign changes of {\em revised sign list} $v=2$, so the number of distinct real roots of $f(x)$, $l-2v$, equals $0$, which means for any $x\in\R$, $f(x)>0$.\par
If $D_4(f)=0$ and $D_3(f)<0$, then $l=3$, the number of the sign changes of {\em revised sign list} $v=2$. Thus, the number of distinct real roots of $f(x)$, $l-2v$, equals $1$, and the number of the pairs of distinct conjugate imaginary root of $f(x)$, $v$, equals $1$, so $f$ has a real root with multiplicity two, which means for any $x\in\R$, $f(x)\ge0$.
\end{pf}
\end{lem}

Now, we can provide a quantifier-free formula of the positive semidefinite cyclic ternary quartic form.
\begin{thm}
Given a cyclic ternary quartic form of real coefficients
$$F(x,y,z)=\sum_{cyc}x^4+k\sum_{cyc}x^2y^2+l\sum_{cyc}x^2yz+m\sum_{cyc}x^3y+n\sum_{cyc}xy^3,$$
then
$$(\forall x,y,z\in \R) \qquad [F(x,y,z)\ge 0]$$
is equivalent to
\begin{align*}
&\vee (g_4=0 \wedge f_2=0\wedge((g_1=0\wedge m\ge1 \wedge m\le4)\vee(g_1>0\wedge g_2\ge0)\vee(g_1>0\wedge g_3\ge0)))\\
&\vee (g_4^2+f_2^2>0\wedge f_1>0 \wedge f_3=0 \wedge f_4\ge0) \\
&\vee (g_4^2+f_2^2>0\wedge f_1>0 \wedge f_3>0 \wedge ((f_5>0 \wedge (f_6<0 \vee f_7<0)) \vee (f_5=0 \wedge f_7<0)))
\end{align*}
where
\begin{align*}
f_1:=&2+k-m-n,f_2:=4k+m+n-8-2l,\\
f_3:=&1+k+m+n+l,f_4:=3(1+k)-m^2-n^2-mn,\\
f_5:=&-4k^3m^2-4k^3n^2-4k^2lm^2+4k^2lmn-4k^2ln^2\\
&-kl^2m^2+4kl^2mn-kl^2n^2+8klm^3+6klm^2n+6klmn^2\\
&+8kln^3-2km^4+10km^3n-3km^2n^2+10kmn^3-2kn^4\\
&+l^3mn-9l^2m^2n-9l^2mn^2+lm^4+13lm^3n-3lm^2n^2\\
&+13lmn^3+ln^4-7m^5-8m^4n-16m^3n^2-16m^2n^3-8mn^4\\
&-7n^5+16k^4+16k^3l-32k^2lm-32k^2ln+12k^2m^2\\
&-48k^2mn+12k^2n^2-4kl^3+4kl^2m+4kl^2n-12klm^2\\
&-60klmn-12kln^2+40km^3+48km^2n+48kmn^2+40kn^3\\
&-l^4+10l^3m+10l^3n-21l^2m^2+12l^2mn-21l^2n^2\\
&+10lm^3+48lm^2n+48lmn^2+10ln^3-17m^4-14m^3n\\
&-21m^2n^2-14mn^3-17n^4-16k^3+32k^2l-48k^2m\\
&-48k^2n+80kl^2-48klm-48kln+96km^2+48kmn+96kn^2\\
&-24l^3-24l^2m-24l^2n+24lm^2-24lmn+24ln^2-16m^3\\
&-48m^2n-48mn^2-16n^3-96k^2-64kl+64km+64kn+96l^2\\
&-32lm-32ln-16m^2-32mn-16n^2+64k-128l+64m+64n+128,\\
f_6:=&4k^2+2kl-4km-4kn+l^2-7lm-7ln+13m^2-mn+13n^2\\
&-40k+20l+8m+8n-32,
\end{align*}

\begin{align*}
f_7:=&-768+352k^2-332l^2+180n^2+180m^2+56k^3-8k^4\\
&+14l^3+132n^3+132m^3+42n^4+42m^4-480k-60lmn-192n\\
&+32klmn-192m+912l+l^4-354kmn+158kln+158klm+26k^2mn\\
&-11kln^2+22k^2lm+22k^2ln-45kmn^2-90lm^2n-45km^2n\\
&-11klm^2+23l^2mn-90lmn^2+kl^2m+kl^2n+36mn-480km+592kl\\
&-480kn-60lm-60ln+8k^3m+8k^3n-20k^2l+32k^2n+32k^2m\\
&-12k^3l+234mn^2+234m^2n-192ln^2-258kn^2-192lm^2-258km^2\\
&+116l^2m+116l^2n+87m^3n+87mn^3-15kn^3+90m^2n^2-30ln^3\\
&-15km^3-30lm^3+25l^2m^2+25l^2n^2-14k^2m^2-14k^2n^2\\
&-146kl^2-10l^3m-10l^3n-2k^2l^2+3kl^3,\\
g_1:=&k-2m+2,g_2:=4k-m^2-8,g_3:=8+m-2k,g_4=m-n.
\end{align*}
\begin{pf}
  By Theorem \ref{thm:eqv}, it suffices to find a quantifier-free formula of
  \begin{align*}(\forall t\in \R)[g(t):=&3(2+k-m-n)t^4+3(4+m+n-l)t^2+k+1+m+n+l-\\
  &\sqrt{27(m-n)^2+{(4k+m+n-8-2l)}^2}t^3\ge0].
  \end{align*}
 Case $1$ $\sqrt{27(m-n)^2+{(4k+m+n-8-2l)}^2}=0$, that is $m=n$ and $4k+m+n-8-2l=0$. Hence
 \begin{align*}
  g(t)=&3(2+k-2m)t^4+3(4+2m-l)t^2+k+1+2m+l\\
  =&3(2+k-2m)t^4+3(8+m-2k)t^2+3(k+m-1).
 \end{align*}
 If $2+k-2m=0$, then $$\forall t\in \R \quad g(t)\ge0 \Longleftrightarrow 1\le m \le 4.$$
  If $2+k-2m>0$, then
 $$\forall t\in \R \quad g(t)\ge0 \Longleftrightarrow (g_1>0\wedge g_2\ge0)\vee(g_1>0\wedge g_3\ge0).$$
Case $2$ $\sqrt{27(m-n)^2+{(4k+m+n-8-2l)}^2}\neq0$ and $1+k+m+n+l=0$. In this case, it is easy to show that $2+k-m-n>0$. Thus,
\begin{align*}
\forall t\in \R,  \quad g(t)\ge0 \Longleftrightarrow&  \forall t\in \R , \quad 3(2+k-m-n)t^2+3(4+m+n-l)\\
&-\sqrt{27(m-n)^2+{(4k+m+n-8-2l)}^2}t\ge0\\
 \Longleftrightarrow& 27(m-n)^2+{(4k+m+n-8-2l)}^2\le 36(2+k-m-n)(4+m+n-l)\\
 \Longleftrightarrow&3(1+k)\ge m^2+n^2+mn.
\end{align*}
Case $3$ $\sqrt{27(m-n)^2+{(4k+m+n-8-2l)}^2}\neq0$ and $1+k+m+n+l\neq0$. In this case, by Lemma \ref{lem:3}, we know that for all $x\in \R$, $g\ge0$ holds if and only if
$$f_1>0 \wedge f_3>0 \wedge ((f_5>0 \wedge (f_6\le0 \vee f_7\le0)) \vee (f_5=0 \wedge f_7<0)).$$
To summarize, the theorem is proved.
\end{pf}
\end{thm}

\begin{ack}
This research was partly supported by President's Fund for Undergraduate Students of Peking University, NSFC-11271034 and the project SYSKF1207 from ISCAS. The author would like to thank the anonymous referees for their valuable comments on a previous version of this paper.
\end{ack}


\begin{thebibliography}{}
\bibitem[ACM84a]{ACM84a} D. S. Arnon, G. E. Collins and S. McCallum: Cylindrical algebraic decomposition
I: The basic algorithm. SIAM J. Comput. 13 (1984): 865--877.
\bibitem[ACM84b]{ACM84b} D. S. Arnon, G. E. Collins and S. McCallum: Cylindrical algebraic decomposition
II: An adjacency algorithm for the plane. SIAM J. Comput.
13 (1984): 878--889.
\bibitem[ACM88]{ACM88} D. S. Arnon, G. E. Collins and S. McCallum: Cylindrical algebraic decomposition
III: An adjacency algorithm for three-dimensional space. J.
Symb. Comput.: 163--187, 1988.
\bibitem[AM88] {AM88}D. S. Arnon and M. Mignotte: On mechanical quantifier elimination for
elementary algebra and geometry. J. Symb. Comput., 5:237--260, 1988.
\bibitem[Br01a] {Br01a}C. W. Brown: Simple CAD construction and its applications. J. Symb.
Comput. 31 (2001): 521--547.
\bibitem[Br01b]{Br01b} C. W. Brown: Improved projection for cylindrical algebraic decomposition.
J. Symb. Comput. 32 (2001): 447--465.
\bibitem[Br12]{Br12} C. W. Brown: Fast simplifications for Tarski formulas based on monomial inequalities. J. Symb. Comput., 7: 859--882, 2012.
\bibitem[BM05]{BM05} C. W. Brown and S. McCallum: On Using Bi-equational Constraints in
CAD Construction. In: Proc. ISSAC2005 (Kauers, M. ed.), 76--83, ACM
Press, New York (2005).
\bibitem[CDMMXX10]{CDMMXX10}C. Chen, J. H. Davenport, J. P. May, M. Moreno Maza, B. Xia, R. Xiao, Triangular Decomposition of Semi-algebraic Systems. Proceedings of ISSAC 2010, ACM Press, 2010.
\bibitem[CJ98]{CJ98} B. F. Caviness and J. R. Johnson(eds.), Quantifier Elimination and
Cylindrical Algebraic Decomposition, Springer-Verlag, 1998.
\bibitem[CH91]{CH91} G. E. Collins and H. Hong: Partial cylindrical algebraic decomposition
for quantifier elimination. J. Symb. Comput. 12: 299--328 (1991).
 \bibitem[CLR87]{CLR87}Choi M D, Lam T Y and Reznick B. Even symmetric sextics, Mathematische Zeitschrift 1987, 195:559--580.
\bibitem[Co75]{Co75} G. E. Collins: Quantifier elimination for real closed fields by cylindrical
algebraic decomposition. In: Lecture Notes in Computer Science 33, 134--165. Springer-Verlag, Berlin Heidelberg (1975).
\bibitem[Co98]{Co98} G. E. Collins: Quantifier elimination by cylindrical algebraic decomposition
- 20 years of progress. In: Quantifier Elimination and Cylindrical
Algebraic Decomposition (Caviness, B. and Johnson, J. eds.), 8--23.
Springer-Verlag, New York (1998).
\bibitem[GLRR89]{GLRR89} L. Gonz\'{a}lez-Vega, H. Lombardi, T. Recio, M.F. Roy: Sturm-Habicht sequence. In Proc.of ISSAC¡¯89, ACM Press, 136--146(1989).
\bibitem[Gon98]{Gon98} L. Gonz\'{a}lez-Vega: A Combinatorial Algorithm Solving Some Quantifier Elimination Problems. In: Quantifier Elimination and Cylindrical Algebraic Decomposition (Caviness, B. and Johnson, J. eds.), 365--375. Springer-Verlag, New York (1998).
    \bibitem[Han11]{Han11}Han J J. An Introduction to the Proving of Elementary Inequalities. Harbin: Harbin Institute of Technology Press, 234--266, 2011 (in Chinese).
\bibitem[Ha99]{Ha99}Harris W R. Real even symmetric ternary forms, Journal of Algebra, 1999, 222: 204--245.
\bibitem[Hilbert88]{Hilbert}Hilbert D. \"{U}ber die Darstellung definiter Formen als Summe von Formenquadraten. Math Ann, 1888, 32: 342--350.
\bibitem[Hong90]{Hong90} H. Hong: An improvement of the projection operator in cylindrical algebraic
decomposition. In: Proceedings of ISSAC ¡¯90 (Watanabe, S. and
Nagata, M. eds.), 261--264. ACM Press, New York (1990).
\bibitem[Hong92]{Hong92} H. Hong: Simple solution formula construction in cylindrical algebraic
decomposition based quantifier elimination. In: Proceedings of ISSAC '92
(Wang, P. S., ed.), 177--188. ACM Press, New York (1992).
\bibitem[La88]{La88} D. Lazard: Quantifier elimination: optimal solution for two classical examples.
J. Symb. Comput. 5:261--266, 1988.
\bibitem[Mc88]{Mc88} S. McCallum: An improved projection operation for cylindrical algebraic
decomposition of three-dimensional space. J. Symb. Comput.: 141--161(1988).
\bibitem[Mc98]{Mc98}S. McCallum: An improved projection operator for cylindrical algebraic
decomposition. In: Quantifier Elimination and Cylindrical Algebraic Decomposition
(Caviness, B. and Johnson, J. eds.), 242--268. Springer-
Verlag, New York (1998).
\bibitem[MB09]{MB09}S. McCallum, C. W. Brown: On delineability of varieties in CAD-based quantifier elimination with two equational constraints. In Proceedings of ISSAC' 2009: 71--78.
\bibitem[Ta48]{Ta48}Tarski A, A decision method for elementary algebra and geometry.
SantaMonica: The RAND Corporation,1948.
\bibitem[Ti03]{Ti03}Timofte V. On the positivity of symmetric polynomial functions, Part I: General results, Journal of Mathematical Analysis and Application, 2003, 284:174--190.
    \bibitem[Ti05]{Ti05}Timofte V. On the positivity of symmetric polynomial functions, Part 2: Lattice general results and positivity criteria for degree 4 and 5. Journal of Mathematical Analysis and Application, 2005, 304:652--667.
\bibitem[Wei94]{Wei94} V. Weispfenning: Quantifier elimination for real algebra - the cubic case.
In Proc. ISSAC 94, Oxford, 1994, ACM Press, 258--263.
\bibitem[Wei98]{Wei98} V. Weispfenning: A New Approach to Quantifier Elimination for Real Algebra.
In: Quantifier Elimination and Cylindrical Algebraic Decomposition
(Caviness, B. and Johnson, J. eds.), 376--392. Springer-Verlag, New York
(1998).
\bibitem[Yang99] {Yang99a}L. Yang: Recent advances on determining the number of real roots of
parametric polynomials. J. Symbolic Computation, 28:225--242, 1999.

    \bibitem[YHZ96]{YHZ96}L. Yang, X. Hou and Z. Zeng: A complete discrimination system for
polynomials. Science in China (Ser. E), 39:628--646, 1996.
\bibitem[YF08]{YF08}Yao Y, Feng Y. Automated decision of positive of symmetric quintic forms. Journal of Systems Scince and Mathematical Sciences, 2008, 28(3):313--324(in
Chinese).
\end{thebibliography}
\end{document}